\documentclass[12pt]{article}

\usepackage{amsmath}
\usepackage{amsfonts}
\usepackage{amssymb}
\usepackage{axodraw}

\def\be{\begin{equation}}
\def\ee{\end{equation}}

\newcommand{\GeV}{{\rm GeV}}
\newcommand{\TeV}{{\rm TeV}}
\newcommand{\eV}{{\rm eV}}

\begin{document}

%%%%%%%%%%
%%%%%%%%%%      title page
%%%%%%%%%%

\begin{flushright} {\footnotesize BICOCCA-FT-02-16}  \end{flushright}
\vspace{5mm}
\vspace{0.5cm}
\begin{center}

\def\thefootnote{\fnsymbol{footnote}}

{\Large \bf Holographic evolution of gauge couplings} \\[1cm]
{\large R. Contino$^1$, P. Creminelli$^1$, E. Trincherini$^2$}
\\[1.5cm]

{\small 
$^1$\textit{Scuola Normale Superiore, Piazza dei Cavalieri 7, I-56126 Pisa,
Italy $\&$ INFN} 
\\[0.3cm]
$^2$\textit{Physics Department, University of Milano-Bicocca, P.zza della Scienza 3,
I-20126 Milano, Italy $\&$ INFN} }

\end{center}

\vspace{1cm}

\hrule \vspace{0.3cm} 
{\small  \noindent \textbf{Abstract} \\[0.3cm]
\noindent
We study the gauge coupling evolution of a unified theory in the compact Randall-Sundrum model with 
gauge bosons propagating in the bulk. One-loop corrections in AdS are interpreted in the 4d 
dual theory as the sum of two contributions: CFT insertions subleading in a $1/N$ expansion and loops of 
the additional particles coupled to the CFT. We have calculated the scalar loop correction to the
low energy gauge couplings both in scenarios where the GUT symmetry is broken by boundary conditions 
and with the Higgs mechanism. In each case our results are what expected from the holographic dual 
theory.

\vspace{0.5cm}  \hrule

\def\thefootnote{\arabic{footnote}}
\setcounter{footnote}{0}

%%%%%%%%%%
%%%%%%%%%%      main part 
%%%%%%%%%%

\section{Introduction}

Grand Unified Theories (GUT) with extra dimensions can address some of the longstanding problems 
of their 4-dimensional (4d) counterparts, while maintaining their virtues.
The doublet-triplet splitting problem, for example, is elegantly solved in models where
the GUT symmetry is broken by the boundary conditions of the gauge fields in the extra dimensions 
and not through a Higgs mechanism~\cite{Kawamura:1999nj,Altarelli:2001qj}. 
The more and more stringent limits on proton decays~\cite{Ganezer:qk}, which are recently 
getting the minimal 4d $SU(5)$ supersymmetric model into trouble~\cite{Murayama:2001ur},
are satisfied in 5-dimensional (5d) extensions~\cite{Hall:2001pg}. 

Much attention has been devoted to the case of flat extra dimensions. 
Here physics appears 4-dimensional in every experiment with typical energies below
the inverse radius scale $1/R$, and therefore the running of gauge coupling constants
is logarithmic as usual.
At higher energies however, nature becomes truly extra dimensional 
and gauge couplings increase with energy following a power law. 
By simple dimensional analysis, the logarithmic running comes from the evolution of
boundary operators and is associated with a logarithmic divergence, while  
the power law dependence on external momentum reflects a power divergence
in the bulk gauge kinetic term.
A power law increase with energy of the coupling constants seems an attractive way 
to obtain unification of the elementary forces at much lower scales than 
the usual GUT models~\cite{Dienes:1998vh}. 
Even more attractive if one speculates on scenarios with
quantum gravity at the TeV, in which case unification of strong and weak interactions 
with gravity seems a realistic ambition.
Unfortunately, even if one postulates a sufficient separation of scales between $1/R$
and the cutoff $\Lambda$ to have an extra-dimensional field theory regime,  
power law evolution comes together with power threshold corrections, which represent the
dominant effect, spoiling completely the predictability. The situation is even worse if one demands unification
to occur at the cutoff scale, because this is right the energy domain in which 
effective field theory breaks down and perturbation theory becomes unreliable.
Therefore, if one is so ambitious to insist on predictive schemes, unification has to 
occur as a result of the slow logarithmic running.
This means that the paradigm of a desert between the electroweak scale and a 
high energy GUT scale $1/R\sim 10^{16}$ GeV, is still valid.
A very useful tool in this case is the effective field theory 
approach of Weinberg~\cite{Weinberg:1980wa}: a matching is performed at energy $\mu\sim 1/R$,
between the full extra-dimensional GUT theory and a 4d theory with only the Standard Model (SM) 
gauge degrees of freedom.
The heavy GUT states, which have masses of the order of $1/R$, are
integrated out and contribute with calculable threshold corrections~\cite{Contino:2001si}.

The picture drastically changes if we depart from the assumption of flatness.
A very interesting situation is the case of just one compact extra dimension, 
in which the metric is that of Anti-deSitter (AdS) space: the Randall-Sundrum
(RSI) model~\cite{Randall:1999ee}.
This model has been originally constructed to address the hierarchy problem, 
with only gravity propagating in the bulk and the SM confined on the TeV brane.
However, its many surprising features have led many groups to explore the possibility of 
realizing a GUT theory in this warped geometry, with gauge bosons propagating in the 
bulk~\cite{Pomarol:2000hp}-\cite{Agashe:2002bx}.
In RSI, physics appears 5-dimensional at energies higher than the AdS curvature $k$ (which is
taken to be of the order of the Planck scale), when everything goes as in the flat limit.
Below this scale and down to the weak scale, there is a huge range of energies in which 
the model is conjectured to be dual to a 4d conformal field 
theory~\cite{susy}, 
along the prescription of the AdS/CFT correspondence~\cite{Maldacena:1997re,Gubser:1998bc,Witten:1998qj}.
This duality allows us to infer that gauge couplings run logarithmically until very high 
energy, even if new GUT physics, namely the Kaluza-Klein (KK) resonances of the unified gauge bosons, 
appears at the TeV scale revealing the unified character of the fundamental forces~\cite{Pomarol:2000hp}.
No surprise then that an effective theory description a l\`a Weinberg 
does not exist beyond the TeV: changing the GUT group and its breaking modifies the 
properties of the Conformal Field Theory (CFT) and consequently the evolution of 
gauge couplings until unification, not only some minor threshold corrections.

In~\cite{Randall:2001gb} the one-loop correction to the low energy coupling was
computed in RSI for a non-abelian gauge theory, employing a momentum cutoff which depends on the fifth
dimension. In~\cite{Goldberger:2002cz}, using dimensional regularization, the case of massless scalar 
QED was considered, while in~\cite{Agashe:2002bx} also the massive case has been studied, adopting 
a Pauli-Villars regulator.
In this work, we further study the scalar QED case with the computation of the gauge field zero-mode 
propagator in 5d for different choices of boundary conditions and for a generic scalar bulk mass. In doing that, 
we choose dimensional regularization that, we believe, is the most economical 
and transparent regulator which preserves the symmetries of the AdS background.
All our results are compatible with what the holographic duality requires. 
This is sufficient to discuss the scalar contribution to gauge coupling evolution
in different GUT scenarios where the unified group is broken by the boundary conditions
or through a Higgs mechanism. 

Section \ref{brane} is devoted to understand the meaning of the evolution of gauge couplings 
in AdS space. Planck brane correlators are the only meaningful observables at energies higher than 
the TeV scale, while mode-by-mode quantities become strongly coupled. We explicitly show how to interpret 
loop corrections to these observables from the holographic point of view. 
In section \ref{low} we present the calculation of the scalar loop correction to the low
energy gauge couplings, for generic boundary conditions and mass. We leave all the computational details 
to the appendix. We use these results in section \ref{GUT} to discuss various mechanisms of GUT symmetry
breaking pointing out the agreement with the holographic interpretation. Conclusions
are drawn in section \ref{conclusions}, where we comment on possible phenomenological scenarios.

\section{\label{brane}Holographic interpretation of the running}
The Randall-Sundrum model with two branes \cite{Randall:1999ee} is simply given by a slice of 
AdS space with metric
\be
\label{eq:metric}
d s^2 =\frac{L^2}{z^2} (dx^\mu dx^\nu \eta_{\mu\nu} -dz^2) \qquad\qquad k \equiv 1/L\;,
\end{equation}
bordered by two branes respectively at $z= z_0 = 1/k$ (Planck brane) and at $z = z_1 \sim \TeV^{-1}$ (TeV brane). The 
attractive feature of the model is the possibility to solve the hierarchy problem through
a gravitational red-shift of energy scales. Many puzzles of this geometry are better understood
through the AdS/CFT correspondence \cite{Maldacena:1997re,Gubser:1998bc,Witten:1998qj}:
the holographic dual of this model \cite{susy} is a quasi-conformal,
strongly coupled 4d theory coupled to 4d gravity. The TeV brane describes a spontaneous breaking
of the conformal invariance \cite{Arkani-Hamed:2000ds,Rattazzi:2000hs}, so that any field living on 
it is seen, in the dual picture, as a bound state of the CFT itself. Also all the KK modes are seen 
as CFT condensates, similarly to the resonances of QCD around the GeV scale. 

In the following, we 
consider GUT models where the gauge bosons of the unified group propagate in the AdS bulk
and study how this reflects on the low energy gauge couplings.   
At energies much greater than the TeV scale, the KK states become strongly coupled. Nevertheless,
if we restrict to the study of inclusive quantities, given by Green functions on the Planck brane,
we can reach energies as high as the Planck scale without entering a strong coupling regime 
\cite{Arkani-Hamed:2000ds,Rattazzi:2000hs} (see also~\cite{Goldberger:2002cz}). 
This is possible because of the exponential die-off of the 
propagators in the bulk, $G \sim e^{-\sqrt{p^2} z}$ at distances $z \gtrsim p^{-1}$, which makes the 
high energy processes on the Planck brane insensible of what is going on deep inside AdS: the 
local cut-off for an observer living on the
Planck brane is given by the AdS curvature $k$. The importance of these inclusive quantities is clear 
also from the holographic point of view, having the brane-brane correlators a simple 4-dimensional meaning. 
In our case, the gauge propagator between two points on the Planck 
brane tells us the strength of the gauge interaction in the 4-dimensional dual theory and it remains 
perturbative despite the fact that the KK gauge bosons become strongly coupled above the TeV.
This dual picture allows us to understand why the gauge coupling running is still logarithmic above the
TeV scale: the CFT composites become broader and broader and the true degrees of freedom emerge, but their 
contribution to the running still remain perturbative and 4-dimensional, {\em i.e.}~logarithmic.
In a unified model, brane-brane gauge correlators for different groups are the same much 
above the unification scale and this may happen in a regime ($E \gg \TeV$) in which only boundary 
correlators make sense. 
 
At energies much greater than the TeV scale, but smaller than the AdS curvature $k$, 
the tree level Planck brane-brane gauge propagator is given by \cite{Pomarol:2000hp}
\be
\label{eq:treeprop}
G(q) = \frac{g_5^2/L}{q^2 (\log(2k/q) -\gamma)} \;,
\end{equation}
where $\gamma$ is the Euler-Mascheroni constant. The holographic interpretation of this formula is 
quite simple \cite{Arkani-Hamed:2000ds}. It just describes the corrections to a 4d vector propagator 
given by $\langle J J \rangle_{\rm CFT}$ insertions, where $J$ is the CFT current coupled to the gauge 
boson (see figure~\ref{fig:tree}). 
Conformal invariance tells us that $\langle J(p) J(-p)\rangle \propto p^2 \log p^2$, so that the logarithmic 
running (\ref{eq:treeprop}) follows. It is worthwhile noting that this CFT running is simply described 
by the tree level AdS propagator and it is common to any gauge group. It follows that, whatever GUT symmetry 
breaking mechanism we choose, {\em the leading CFT contribution to the running is always GUT invariant}.
From eq.~(\ref{eq:treeprop}) we see that the CFT gives a positive contribution to the beta-function
$b_{\rm CFT} = 8 \pi^2 L /g_5^2 \sim N^2$, where $N$ is the number of colors of the conformal theory
\cite{Freedman:1998tz}. This should be large to ensure that the non-renormalizable 5d
gauge theory makes sense: the AdS curvature $k$ must be much smaller than the cut-off scale 
$\Lambda = 24 \pi^3/{g_5^2}$
or, equivalently, the number $N$ of colors should be large.

%%
%% Tree Level
%%
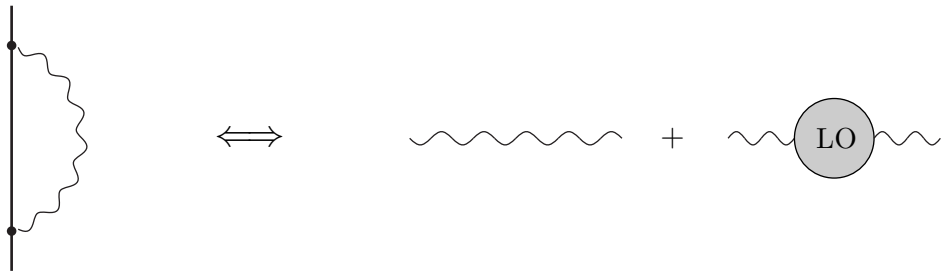
\begin{figure}[t]\begin{center}
\begin{picture}(380,100)
\SetWidth{1} \Line(10,0)(10,100) \SetWidth{0.5}
\Vertex(10,15){1.8} \Vertex(10,85){1.8} \PhotonArc(0,50)(36.5,-70,70){2}{7}
\Text(100,50)[]{\large $\Longleftrightarrow$}
\Photon(160,50)(240,50){-2.5}{5} \Text(260,50)[]{+}
\Photon(280,50)(305,50){2.5}{2} \GCirc(320,50){15}{0.8}
\Text(322,50)[]{\small LO} \Photon(335,50)(360,50){2.5}{2}
\end{picture}
\caption{\label{fig:tree} \small The brane-brane correlator in AdS corresponds holographically
to the free gauge propagator corrected by the LO contribution in $1/N$ of the CFT (of order 
$\sim {\cal {O}}[N^2 (\alpha/4\pi)]$ with respect to the tree level).
The grey circle represents the $\langle JJ \rangle$ insertion.}
\end{center} \end{figure}
\vspace{0.3cm}

\subsection{Radiative corrections to brane correlators}
Additional contributions to the running of the gauge couplings come from loop corrections to the brane-brane
propagator. It is therefore natural to ask what is the holographic interpretation of these loops.  
For example, what is the 4-dimensional counterpart of the vacuum polarization due to a bulk scalar? 
In the limit in which we remove the Planck brane, obtaining a complete AdS space, we know that the dual picture 
is simply a CFT. In this case, bulk loops are interpreted as corrections to the CFT correlators, subleading in 
a $1/N$ expansion \cite{Aharony:1999ti}. Concerning the scalar loop correction to the 
gauge two-point function, we would find a modification of the $\langle J J\rangle$ CFT correlator. As the 
dependence of this correlator on the 4d momentum is fixed by conformal invariance, only the coefficient in 
front receives $1/N$ corrections.

What changes if we add the Planck brane? The rough picture is the following. Cutting off 
the part of AdS space near its boundary corresponds to a UV modification of the CFT, which is
now smeared over a distance of order $k^{-1}$: degrees of freedom of 
shorter wavelength have been integrated out. Moreover the 4d role of fields living in AdS space 
changes. In the full AdS case they are not dynamical from the 4d point of view: their boundary 
behaviour at infinity just acts as a source for the corresponding operator of the CFT. With 
the addition of the Planck brane, bulk fields become dynamical also from the 4d viewpoint, as we must
integrate over their boundary value on the brane. 

We thus expect that radiative corrections to brane correlators in presence of the Planck brane 
describe not only $1/N$ subleading CFT terms, but the additional contribution of the 4d fields
made dynamical by the introduction of the brane. If we have a scalar field in AdS cut by the 
Planck brane, the 4d theory contains a dynamical scalar, coupled to the CFT through an operator $O(x)$, 
which has dimension 4 if the scalar is massless. Loops of this 4d scalar will enter the running 
of the gauge couplings. 

%%
%% OneLoop
%%
\begin{figure}\begin{center}
\begin{picture}(380,100)
\SetWidth{1} \Line(10,0)(10,100) \SetWidth{0.5} \Vertex(10,15){1.8}
\Vertex(10,85){1.8}
\PhotonArc(0,50)(36.5,-70,70){2}{7} \BCirc(36.5,50){15}
\Text(100,50)[]{\large $\Longleftrightarrow$}
\Photon(160,75)(185,75){2.5}{2} \CArc(200,75)(15,0,360)
\Photon(215,75)(240,75){2.5}{2}
\Text(260,75)[]{+}
\Photon(280,75)(305,75){2.5}{2} \GCirc(320,75){15}{0.8}
\Text(321,75)[c]{\small NLO} \Photon(335,75)(360,75){2.5}{2}
\Text(200,103)[c]{\small (a)} \Text(320,103)[c]{\small (b)} \Text(380,75)[]{+}
\Photon(160,25)(185,25){2.5}{2} \CArc(200,25)(15,0,360)
\GBoxc(200,40)(10,10){0.8}  \Photon(215,25)(240,25){2.5}{2}
\Text(200,-3)[c]{\small (c)}
\end{picture} \vspace{0.5cm}
\caption{\label{fig:loop}\small The one-loop (rainbow) scalar correction to the brane-brane correlator in
AdS corresponds holographically to three different diagrams: a 4d scalar loop graph (a), the
same diagram with the scalar propagator corrected by the CFT (c), and the NLO contribution
in $1/N$ of the CFT (b). Diagrams (a), (b) are both ${\cal {O}}(\alpha/4\pi)$ with respect to the
tree level; diagram (c) is negligible because the scalar coupling to the CFT is $M_{\text{Pl}}$-suppressed. 
The grey circle (square) represents the $\langle JJ \rangle$ ($\langle OO \rangle$) insertion. A
similar holographic interpretation holds for the seagull diagram.}
\end{center} \end{figure}
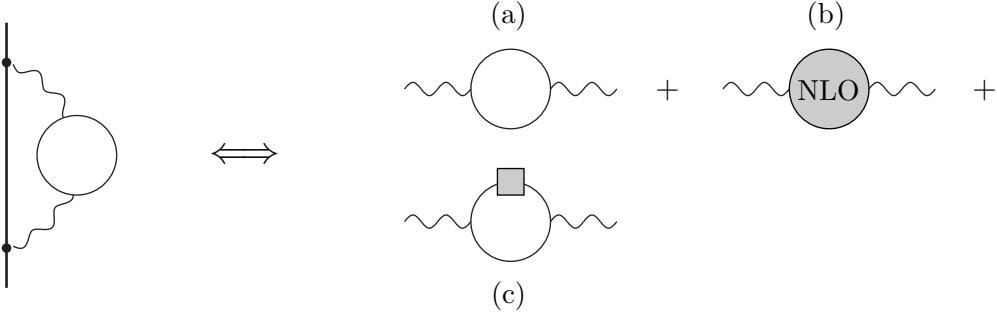

As depicted in figure \ref{fig:loop}, the one-loop AdS correction corresponds to the sum of different 
terms:  the contribution from the 4d scalar (a), whose propagator gets itself a CFT correction (c), 
and the NLO CFT insertion (b). It is worth noting that the various terms can be arranged in a double
expansion: the first is the standard series in powers of $(\alpha/4\pi)$, the second 
is the expansion of the CFT correlators in powers of $1/N$. The two expansions are related, as the 
holographic prescription tells us that $1/N^2 \sim g_5^2/16 \pi^2 L$. Diagrams (a) and (b) 
are of order ${\cal O}(\alpha/4\pi)$ with respect to the tree level; diagram (c) is completely 
negligible in this case, being the CFT coupled to the 4d scalar only through $M_{\rm Pl}$-suppressed
operators. The corresponding diagram in the case of vector boson loops is ${\cal O}[N^2 (\alpha/4\pi)^2]$,
but still subleading with respect to the other two contributions, as 4d perturbativity requires 
$N^2 (\alpha/4\pi) \ll 1$.

We can look at the contribution (b) and (a) in fig.~\ref{fig:loop} as coming respectively from
the limiting case of a 5d loop deep inside AdS or close to the Planck brane. This is quite
intuitive, as the 4d scalar field comes from the integration over the boundary conditions on the Planck 
brane. In the complete AdS case, the boundary values $\phi_0, A_\mu^0, g_{\mu \nu}^0$ for the
various fields at infinity act as sources for the corresponding operators in the CFT \cite{Witten:1998qj}: 

\be
\label{eq:AdSCFT}
\langle e^{-\int d^4 x\; O(x) \phi^0(x) + J^\mu(x) A_\mu^0(x) + T^{\mu\nu}(x) g_{\mu\nu}^0(x) }\rangle_{\rm 
CFT} = e^{-S_{\rm AdS}(\phi^0, A_\mu^0, g_{\mu\nu}^0)} \;.
\end{equation}

The right hand side of this equation must be regularized \cite{Witten:1998qj}, and this procedure 
leads us closer to the truncated AdS case we are interested in. The standard procedure is to limit the 
$z$ integration to $z > \epsilon$ (which corresponds to introducing an explicit UV cut-off on the CFT), 
add a proper local counterterm action (divergent for $\epsilon \rightarrow 0$) function of $\phi^0, 
A_\mu^0, g_{\mu \nu}^0$ and their derivatives, and then take the limit $\epsilon \rightarrow 0$. 
In the case with only a scalar field, eq.~(\ref{eq:AdSCFT}) becomes 

\be
\label{eq:AdSCFTreg}
\langle e^{-\int d^4 x\; O(x) \phi^0(x)}\rangle_{\rm CFT}
= \lim_{\epsilon \rightarrow 0} e^{-S_{\rm AdS}(\phi^0, \epsilon)} e^{-S_{\rm count}(\phi^0,\epsilon)} \;.
\end{equation}    

Suppose now not to perform the final limit, keeping an explicitly truncated AdS space. As we have 
integrated out a portion of space which corresponds to the UV of the CFT, we expect this to correspond 
to a smearing procedure in which fast modes are integrated out~\footnote{The counterterm action contains 
an infinite series of increasing dimension, properly suppressed by powers of
$k$: at energies of order of the AdS curvature the theory becomes non-local. In the
following we concentrate on the lower dimension operator $(\partial \phi_0)^2$.} 
\cite{susy,Balasubramanian:1999jd,Perez-Victoria:2001pa}. 
At this stage, the scalar loop correction in AdS of figure \ref{fig:loop} gives a subleading 
contribution to the $\langle J J\rangle_{\phi_0}$ CFT correlator in the external background $\phi_0$.

The last step to get the Randall-Sundrum scenario is to integrate over the boundary values 
$\phi^0, A_\mu^0, g_{\mu \nu}^0$, which become dynamical fields, introducing a generic brane action 
$S_{\rm bound}(\phi_0)$. Consider for instance a brane action with only a kinetic term proportional
to an arbitrary parameter $\xi$:
\be
\label{eq:xi}
S_{\rm bound}(\phi_0) = \frac{\xi}{k} \int_{\rm brane} \!\!\!\!\! d^4x \sqrt{g} \, \partial_\mu 
\phi \partial_\nu \phi^\dagger  g^{\mu\nu} \;.
\end{equation} 
By varying $\xi$ one changes the kinetic term of the 4d scalar and therefore the relative importance 
between its loop contribution (fig.~\ref{fig:loop}a) and the CFT correction 
(fig.~\ref{fig:loop}b)~\footnote{Note however that, if the integration over the boundary conditions 
were done after the addition 
of the counterterm $S_{\rm count}(\phi_0)$ (eq.~(\ref{eq:AdSCFTreg})), the theory would be ill-defined. 
This is expected because, in this case, the dependence of the 4d action on $\phi_0$ is given by the two 
point function $\langle O O \rangle_{\rm CFT}$ of the corresponding operator \cite{Witten:1998qj}: 
$S(\phi^0) \propto \int d^4x d^4x' \phi^0(x) \phi^0(x')/|x-x'|^8$.
This implies that, if the field $\phi_0$ is seen as dynamical, it has a non-local kinetic term of the
form $q^4 \log q$. This property can be used to find $S_{\rm count}(\phi_0)$ at leading order. 
To fix it we {\em require} that the 5d brane-brane propagator goes as $1/(q^4 \log q)$. It is easy
to obtain that this happens for $\xi = -1$. For $\xi > -1$ the integration over the boundary conditions
is well defined as the 4d scalar has a conventional kinetic term, with the correct sign.}.
In the limit $\xi \to +\infty$, the 4d scalar is frozen out and we are left with the CFT correction; the same 
result holds by choosing Dirichlet boundary conditions on the Planck brane.   
From these considerations, it should be clear the strict connection between boundary terms in AdS and
the 4d scalar mode. Moreover, all the features of the AdS bulk reflect on the CFT. In particular,
if the GUT symmetry is unbroken in the bulk, the CFT is GUT-preserving at all orders.

\subsection{The CFT contributions}
We now concentrate on the pure CFT corrections, as if we had pushed the Planck brane to infinity, 
recovering the complete AdS space. We have shown that at leading order the $\langle J J \rangle$
correlator does not distinguish among the unbroken subgroups of a unified theory, while at NLO the CFT correction is 
GUT invariant or not depending on the mechanism we choose to break the GUT symmetry. 
If the symmetry is broken by the boundary conditions, the AdS bulk remains GUT invariant as well as the dual 
CFT \footnote{If the boundary conditions on the TeV brane break the GUT symmetry, in the dual picture 
we have a spontaneous breaking of the symmetry at the TeV scale, together with the conformal breakdown. 
Still, the running above the TeV scale remains GUT invariant, as we will explicitly check in
the following sections.}. 
In this case, at subleading order the CFT still gives a common running to all the unbroken subgroups. 

Another possibility is that the unified theory is broken in the bulk, through a vev of a charged scalar 
$\Sigma$. If the expectation value of the scalar is constant along the fifth dimension, the 
conformal symmetry is still unbroken (all AdS isometries are preserved) but the GUT symmetry is 
not \footnote{We may also consider operators which break the conformal symmetry, corresponding to massive
scalars. In this case the scalar profile will not be constant in AdS.}.
In the holographic theory, we have turned on an operator $O_{\Sigma}$ coupled to the 4d $\Sigma$ scalar, 
transforming under the GUT symmetry, which therefore results spontaneously 
broken~\footnote{In presence of the Planck brane, the GUT symmetry breaking is spontaneous, due to the 
vacuum expectation value of the $\Sigma$ field. In 
the complete AdS case one just turns on the corresponding operator $O_\Sigma$, causing an explicit
breaking.}.  In this scenario, GUT-breaking corrections to Planck brane propagators correspond to analytic or 
non-analytic operators, involving $\Sigma$, in the 5d effective action. 
The contribution due to analytic operators is not calculable, and can be only estimated through a naive dimensional 
analysis. For instance, in the case of the $\Sigma F F$ 
operator, naive dimensional analysis gives a ratio between the ${\cal{O}}(1)$ and ${\cal{O}}(N^2)$ corrections to 
the CFT beta-function:  
\be
\label{eq:NDA}
\frac{b_{\rm CFT}^{\rm NLO}}{b_{\rm CFT}^{\rm LO}} \sim  \frac{g_5 \langle \Sigma \rangle}{\Lambda} 
= \frac{M_{\rm GUT}}{\Lambda}\;,
\end{equation}
where $\Lambda = 24 \pi^3 /g_5^2$ is the 5d cut-off. The second equality relates this
ratio to the mass of the 4d GUT gauge bosons: $M_{\rm GUT} = g_5 \langle \Sigma \rangle$. If 
we want unification to occur in the regime in which the holographic dual makes sense, we have to
require $M_{\rm GUT} \ll k$, so that the ratio (\ref{eq:NDA}) must be quite small. 

From the 4d point of view these corrections to the $\langle J J\rangle_{\rm CFT}$ correlator
are due to the multiple Green functions involving the additional operator $O_\Sigma$:
\be
\label{eq:multiple}
\langle O_\Sigma O_\Sigma \ldots J J\rangle_{\rm CFT} \;.
\end{equation} 
For example the $\Sigma F F$ AdS vertex gives at tree level a CFT correlator $\langle O_\Sigma J 
J\rangle_{\rm CFT}$ \cite{Freedman:1998tz}; turning on the $O_\Sigma$ operator thus modifies the 
current-current correlators in a non GUT-invariant way. All these corrections are suppressed by
powers of $1/N$ and $\lambda \equiv \langle\Sigma\rangle/ k^{3/2}$, where $\lambda$ is the 
coupling constant of the operator $O_\Sigma$ in the 4d picture.

Notice that the non-calculability due to higher-dimension operators in AdS is here 
reflected into an incalculable CFT beta-function. Nevertheless predictability is retained if we assume
that the AdS picture is weakly coupled so that the perturbative expansion makes sense, with higher 
dimension operators properly suppressed by powers of the 5d cut-off $\Lambda$. 
In this case, CFT contributions which distinguish among the unbroken subgroups are suppressed with 
respect to the GUT invariant leading CFT running. In turn, this leading contribution cannot be 
too large if we want to build a phenomenologically viable model.
 
The reason for this is quite simple and it is a general problem of unification in this kind of models: 
the GUT-invariant CFT running would imply, for $N \gg 1$, that {\em at low energy we should see nearly 
SU(5)-invariant couplings: $\Delta \alpha/\alpha \ll 1$}. Alternatively: if $N$ is too large, we meet 
the strong coupling regime before reaching the unification scale. It is easy to deduce a limit on $N$ 
from the requirement of perturbativity at the GUT scale (which we take to be of the order of the standard 
one $M_{\rm GUT} \sim 10^{16} \GeV$): $N^2 ( \alpha_{\rm GUT}/4 \pi) \ll 1$, where the $N^2$ factor 
comes from the number of CFT states. From this we obtain:
\be
\label{eq:perturbative}
b_{\rm CFT} \ll \frac{2\pi \alpha_i^{-1}(\TeV)- b_i^0 \log{M_{\rm GUT}/\TeV}}{1+\log{M_{\rm GUT}/\TeV}} 
\sim 8 \;,
\end{equation}
where the numerical bound is obtained for $b_i^0$ given by the SM matter 
content. An opposite bound on $b_{\rm CFT}$, or equivalently on $N$, comes from the requirement of 
perturbativity in 5d, namely $\Lambda/k\pi\gg 1$, where $\Lambda$ is the 5d cutoff: the inequality 
\be
\label{eq:AdSpert}
b_{\rm CFT} = \frac{8\pi^2 L}{g_5^2} \gg \frac13 
\end{equation}
follows.

The general conclusion is that the leading CFT running cannot be much greater than other contributions
which separate the unbroken subgroups, coming from additional particles coupled to the CFT. The limit on $N$
is not so strong to spoil the perturbativity of the AdS picture, as we see comparing 
eqs.~(\ref{eq:perturbative}) and (\ref{eq:AdSpert}), even if the allowed window is not too wide. This limit on 
the CFT leading contribution implies, in turn, that subleading corrections, coming from bulk loops and 
higher dimension operators are negligible with respect to the non-CFT running. 

In principle we could discuss the gauge coupling running even in absence of a unified group in the bulk and 
check if the gauge couplings cross at a certain energy. In this case the CFT contribution to the running 
of each group is different at leading order, so that the running may be much faster with a consequent 
lowering of the unification scale \cite{Randall:2001gb}. 
However, the CFT beta-function for the three groups, given at leading order by the three independent gauge 
kinetic terms, is incalculable, so that no firm prediction seems possible.

Before moving to the explicit calculations, we want to stress an important conceptual difference between the
standard models of unification and the ones built in AdS space. In the standard case, Weinberg's approach 
of effective gauge theory is very useful and it tells us that the details of GUT-symmetry breaking, resulting
only in threshold corrections, are not
crucial to test unification. Here the situation is different.
Modifying the unified gauge group we are at the same time changing the CFT excitations, hopefully 
around the corner, at the TeV scale. The pattern of symmetry breaking does not influence only the physics
at far-away energies, but also the subleading CFT corrections to the running down to the TeV 
scale. All this follows from the fact that AdS space describes at the same time the CFT properties 
and the behaviour of the additional particles coupled to it.

\section{\label{low}The low energy gauge coupling}

In this section we present our result for the one-loop scalar correction to the low energy coupling
of a $U(1)$ gauge group in the bulk. We leave to the appendix all the computational details, focusing
our attention on the holographic interpretation.
Once given the main formulae for different boundary conditions of the scalar field, we will able in the 
next section to discuss various scenarios of GUT symmetry breaking.

In order to regulate the loop divergence we choose the dimensional regularization, which proved to be 
a powerful scheme also in theories with flat extra-dimensions~\cite{dimreg}.
In the specific case of the one-loop correction to the zero-mode gauge correlator, it is enough
to extend the brane dimension to a generic (complex) value $d$ keeping just one extra 
dimension. 
Analogously to the Minkowski case, the isometries of AdS space are clearly preserved.
The zero-mode gauge self-energy reads, for external 4d momentum $p$:
\begin{equation}
\frac{1}{g^2(p^2)} = \frac{\log (z_1/z_0)}{k\, g_5^2} + \Delta_0(\mu) + \Delta_1(\mu) - \Pi(p^2,\mu) \;,
\end{equation}
where $\mu$ is the subtraction point and $\Delta_{0,1}(\mu)$ are the coefficients of the gauge kinetic 
terms localized on the branes. $\Pi(p^2,\mu)$ is the one-loop scalar correction
\begin{equation}
\label{eq:pisum}
\Pi(p^2,\mu) = - \mu^{4-d} \sum_{\{x_n\}} \int_0^1 \!\!  dx\; (2x-1)^2 \int \frac{d^dq}{(2\pi)^d}
 \, \frac{1}{[q^2+x_n^2+c^2(x)]^2} \;.
\end{equation}
Here $c^2(x)=x(1-x)(-p^2)$ and $x_n$ is the mass of the $n$-th Kaluza-Klein mode of the scalar field
(see appendix). Using the technique described in the appendix, it is easy to perform the integration first 
and then the sum, getting
\begin{equation} \label{eq:pi}
\begin{split}
\Pi(p^2,\mu) = \frac{(b_0/2)}{8\pi^2} \bigg[ & -\frac{\alpha}{\epsilon} +
 \log\left(\sqrt{-p^2}\sqrt{z_0 z_1}\right) + \alpha \log\frac{\sqrt{-p^2}}{\mu} \\
 & + 3 \int_0^1 \!\! dy\; y\sqrt{1-y^2}\; \log f\left(iy\sqrt{-p^2}/2\right) \\
 & + \alpha \frac{\gamma}{2} + \log\pi - \frac{4}{3} (1+\alpha) \bigg] \;.
\end{split}
\end{equation}
With $b_0=1/3$ we mean the beta-function of a charged 4d scalar, and $d=4-\epsilon$.
The previous formula is a completely general result, valid for a scalar with arbitrary boundary
conditions and mass; in the case of $(\pm\pm)$, $(\pm\mp)$ boundary conditions, one should read
$\alpha=\pm 1$, $\alpha=0$ respectively and choose a function $f=f_{\pm\pm}$, $f=f_{\pm\mp}$, whose
expression is given in appendix. In the particular case of a $(++)$ massless scalar, 
eq.~(\ref{eq:pi}) coincides with the result of~\cite{Goldberger:2002cz}.

The zero-mode gauge propagator is an exclusive observable and does not make sense above the TeV
where the 0 mode becomes strongly coupled.
This means that eq.~(\ref{eq:pi}) can be really trusted only for external momenta 
$|p|\lesssim \TeV$~\cite{Goldberger:2002cz};
at these energies it matches the Planck brane-brane correlator, therefore admitting a simple holographic
interpretation. Once the function $f$ in eq.~(\ref{eq:pi}) is expanded 
for $z_1 |p|\ll 1$, the logarithmic dependence on the momentum $p$ must be the correct one 
for an infrared log.
The logarithmic divergence, represented by the $1/\epsilon$ pole, is the same as in the flat limit
(for the latter, see~\cite{Contino:2001si}).
This was expected, because in the very high energy regime the curvature can be neglected and AdS
appears locally flat~\cite{Goldberger:2002cz,Agashe:2002bx}.

In the following we collect the low energy limit $z_1|p|\ll 1$ expression of $\Pi(p^2,\mu^2)$
for all possible choices of boundary conditions in the massless case and for a (++) scalar with 
AdS bulk mass $m$.
Using the asymptotic expansions of eqs.~(\ref{eq:asymptotes}), we obtain (subtracting the $1/\epsilon$ 
divergence and omitting irrelevant constants):

\newpage
\begin{center}  \textsc{massless scalar} \hspace{0.1cm} ($1/z_1\gg |p|> z_0/z_1^2$) \end{center}
\vspace{-0.2cm}
\begin{align}
\Pi_{++}(p^2,\mu) &\simeq \frac{b_0}{8\pi^2} \left[\log\frac{z_1}{z_0}+ \log z_0 \sqrt{-p^2} 
 - \frac{1}{4} \log\mu z_0 - \frac{1}{4} \log\mu z_1  \right] \label{eq:massless1} \\[0.1cm]
\Pi_{--}(p^2,\mu) &\simeq \frac{b_0}{8\pi^2} \left[ \log\frac{z_1}{z_0}+ 
  \frac{1}{4} \log\mu z_0 + \frac{1}{4} \log\mu z_1  \right] \\[0.1cm]
\Pi_{-+}(p^2,\mu) &\simeq \frac{b_0}{8\pi^2}\,\frac{3}{4} \log\frac{z_1}{z_0} \\[0.1cm]
\Pi_{+-}(p^2,\mu) &\simeq \frac{b_0}{8\pi^2} \left[ \frac{5}{4} \log\frac{z_1}{z_0}
 + \log z_0 \sqrt{-p^2}  \right] \;.\label{eq:massless4}
\end{align}
\vspace{0.4cm} 
\begin{center}  \textsc{massive (++) scalar} \hspace{0.1cm} ($k\gg m\gg |p|$, $|p|\ll 1/z_1$) \end{center}
\vspace{0.3cm}
\begin{equation} \label{eq:massive1}
\Pi(p^2,\mu) \simeq \frac{b_0}{8\pi^2} \left[\log\frac{z_1}{z_0}+ \log m z_0 
 - \frac{1}{4} \log\mu z_0 - \frac{1}{4} \log\mu z_1  + \frac{m^2 z_0^2}{8} 
 \left( \log\frac{z_1}{z_0}-\frac{1}{2} \right)  \right] \;.
\end{equation}
\vspace{0.4cm} 
\begin{center}  \textsc{massive (++) scalar} \hspace{0.1cm} ($m\gg k\gg 1/z_1\gg |p|$) \end{center}
\vspace{0.3cm}
\begin{equation} \label{eq:massive2}
\Pi(p^2,\mu) \simeq \frac{b_0}{8\pi^2} \left[ \frac{3}{4}\log\frac{z_1}{z_0} 
 + \frac12 \log\frac{m}{\mu} + \frac12 z_0 \sqrt{m^2} \log\frac{z_1}{z_0} \right] \;.
\end{equation}
\vspace{0.2cm}

\noindent For the $(++)$ scalar these equations agree with the results of~\cite{Agashe:2002bx}
if $\mu=k$.

The $\log p$ terms in the previous formulae are the expected infrared logarithms.
Holographically they correspond, in the $(++)$ massless case, to the 4d massless mode which 
runs logarithmically from high scale down to low energy.
It can be interpreted as the Goldstone boson of the symmetry $\phi\to\phi\, +\,$const., which shifts the 5d
scalar field by a constant.
This symmetry is broken in the $(+-)$ case by the boundary condition on the TeV brane and the
Goldstone boson acquires a tiny mass $M\sim z_0/z_1^2 \sim 10^{-4}\, \eV$.
That $M$ should be so small can be understood by the following argument: the 4d scalar couples to the CFT with a
$M_{\rm Pl}$ suppressed operator and the analog of the pion decay constant is $f_\pi\sim k$.
A Dirichlet boundary condition on the TeV brane implies the breaking of the $\phi\to\phi\, +\,$const.
symmetry with a typical breaking scale $\sim \TeV$. Then a mass follows for the pseudo-Goldstone 
boson $M^2\sim \TeV^4/f_\pi^2\sim \TeV^4/k^2$ (the exact value of the mass can be derived as 
the lightest eigenvalue of the spectrum equation~(\ref{eq:spectrum}) given in appendix 
\footnote{Also ref.~\cite{Huber:2002np}
has recently pointed out the appearance of such a small eigenvalue in the similar case of a boundary 
mass term on the TeV brane for the scalar field.}).
We conclude that, for scalar boundary conditions $(+-)$, the holographic theory contains an almost
massless 4d scalar contributing to the running of the gauge coupling down to very low 
energy \footnote{In the case of 
a vector field, Dirichlet boundary conditions on the TeV brane implies
a mass $\sim \TeV$. This is expected, because its coupling with the CFT is dimensionless so that the
mass is only logarithmically suppressed by $1/\log(k/\TeV)$. For a fermion field we obtain 
$m^2 \sim \TeV^2 \cdot \TeV/k$.}. This explains the $\log p$ term in $\Pi_{+-}$.

Non-analytic operators in the bulk, like $\sqrt{\cal R}FF$, 
in a background with a constant scalar curvature ${\cal R}\propto k^2$ give a bulk $k FF$ operator
which corresponds in $\Pi(p^2,\mu)$ to calculable $\log z_1/z_0$ terms~\footnote{The non-analyticity 
is a consequence of the fact that we need a term linear in k, while the
metric is a function of $k^2$. We thank Riccardo Rattazzi for clarifying us this point.}.
This local bulk effect is interpreted holographically as a calculable correction to the CFT beta-function.
There is also an additional incalculable correction coming from 
the linear divergence of the bulk gauge kinetic term, but of course this does not appear in dimensional
regularization.
Assuming an holographic point of view, one can extract this NLO CFT contribution
from any of the equations~(\ref{eq:massless1})-(\ref{eq:massless4}).
Consider for example eq.~(\ref{eq:massless1}): the $\log pz_0$ term is the running of the 4d scalar
from the Planck scale down to $p$; the latter two terms, coming from the log divergence on the AdS boundary,
are threshold corrections in the 4d theory at the scales $1/z_0$, $1/z_1$.
The remaining contribution, namely the first term in eq.~(\ref{eq:massless1}), is the 
calculable part of the NLO CFT correction.
On the other hand, any of the equations~(\ref{eq:massless1})-(\ref{eq:massless4}) does not
constitutes by itself an unambiguous test of the holographic interpretation.
Such an ambiguity can be resolved only by comparing the different results for the various 
parities as we will do in the next section  when we consider the GUT breaking scenarios. 
In the massive case, eqs.~(\ref{eq:massive1}),(\ref{eq:massive2}), there are also contributions
of the form $z_0^2 m^2\, \log z_1/z_0$, $z_0 \sqrt{m^2}\, \log z_1/z_0$; 
both are calculable, as they correspond to AdS bulk operators which depend non-analytically 
on the scalar curvature ${\cal R}$ (the former) or on the Lagrangian parameter $m^2$ (the latter).
%only the latter ones
%are calculable depending non-analytically on the Lagrangian parameter $m^2$. 
The $z_0 \sqrt{m^2}\, \log z_1/z_0$ terms appear
%This dependence appears 
uniquely for a very large value of the mass, $m \gg k$ (see eq.~(\ref{eq:massive2})). In this limit 
%$z_0 \sqrt{m^2}\, \log z_1/z_0$ terms 
they represent the contribution of CFT operators of very high dimension
$\propto m/k$, while $\log m$ terms can be interpreted only from a 5d point of view:
their coefficient $b_0/2$ comes from the running of boundary operators.

It is well known~\cite{Georgi:2000ks} that, in the 5d flat case, gauge kinetic terms on the boundary
evolve logarithmically with energy, and their beta-function gets a one-loop contribution from particles 
living in the bulk. This evolution is intimately connected with a logarithmic divergence.
The whole tower of massive KK states contributes to the running on a given boundary 
with $1/4$ of the beta-function $b_0$ of the zero-mode, the sign of the effect 
depending on the parity of the loop fields~\cite{Contino:2001si}: including also the zero-mode 
contribution, one finds $\pm 1/4\, b_0$ if the loop field is $\pm$ on that specific boundary.
The logarithmic divergences, together with the associated $\log p/\mu$ terms, cancel 
in the one-loop correction from a $(\pm,\mp)$ scalar to the zero-mode gauge propagator,
summing the contributions at the two boundaries.
All these considerations must remain valid in the warped case as well, being the divergences the same as
in the flat limit. This can be verified looking at eqs.~(\ref{eq:massless1})-(\ref{eq:massive2}).
In the warped case, the contribution from massive KK states to the running of operators 
on the boundaries $z=z_0,z_1$ will freeze out at the typical local scale $1/z_0$, $1/z_1$.
This gives the $\log \mu z_0$, $\log \mu z_1$ terms in $\Pi(p^2,\mu)$ which are the counterpart
of the $\log \mu R$ terms of the flat case. A further source of $\log z_{0,1}$ terms might be 
finite non-local operators which will be in general present in the 5d effective action.
Indeed, the dependence on $z_{0,1}$ of the function $f$ in eq.~(\ref{eq:pi}), is quite complicated before
taking the limit $z_0\ll z_1$. Only when the is a large separation of scales $z_0\ll z_1$, we
recover the simple expression of eqs.~(\ref{eq:massless1})-(\ref{eq:massless4})
required by the holographic interpretation.

Concerning the holographic interpretation of 
the brane kinetic terms in AdS, they correspond to adding a constant term
to the 4d inverse coupling $1/g^2(p^2)$, shifting its Landau pole~\cite{Arkani-Hamed:2000ds}.
In other words, it is a modification of the 4d theory at a scale corresponding to the
position of the brane in AdS.
There is therefore no connection between boundary terms in AdS and log evolution in the holographic
theory. It is remarkable that in the flat case all the logarithmic running comes from boundary operators, 
while the main logarithmic running in the 4d theory dual to RSI comes from the AdS bulk.

\section{\label{GUT}GUT breaking: the holographic point of view}

Armed with the previous results, we discuss now different mechanisms of breaking
the GUT symmetry in AdS, either through suitable boundary conditions for the gauge fields, or
turning on the vev of a scalar field in the bulk.
We consider for simplicity the particular case of an $SU(5)$ group in the bulk broken down to 
$SU(3)\times SU(2)\times U(1)$ and we study the loop correction to the low energy couplings 
given by a scalar multiplet in the fundamental representation.
It is understood that the results have a general validity.

\subsection{GUT breaking through boundary conditions}

Let us consider first the case in which the GUT symmetry is reduced at low energy by
the boundary conditions. We assume that the $SU(3)\times SU(2)\times U(1)$
gauge bosons $A_\mu^a$ have always parity $(++)$,
while the $X,Y$ bosons $A_\mu^{\hat a}$ can be $(\pm,\mp)$ or $(--)$: $SU(5)$ is broken 
on the TeV or Planck brane, or both. 
The relative parities of the doublet and triplet components of the scalar 5-plet $\varphi$ in the bulk 
are fixed by gauge invariance. We choose 
$\varphi_2 =(++)$ for the doublet component and this forces 
$\varphi_3=(\pm,\mp)$, $(--)$ for the triplet when $A_\mu^{\hat a}$ are $(\pm,\mp)$, $(--)$
respectively.

\vspace{0.3 cm}
\noindent \begin{center} \textsc{GUT breaking on the TeV brane} \end{center}
\begin{equation*}
\varphi = \begin{bmatrix} \varphi_2 (++) \\ \varphi_3 (+-) \end{bmatrix}
\qquad\qquad\qquad \text{for} \quad A_\mu^{\hat a} (+-) \;\; A_5^{\hat a} (-+)
\end{equation*}
\vspace{-0.1cm}

A theory with a gauge group $SU(5)$ in pure AdS is dual to a 4d CFT with a {\it global} 
$SU(5)$ invariance. Putting the Planck brane and imposing $+$ conditions for the gauge bosons
corresponds, in the holographic theory, to gauge the global symmetry.
Let us now insert the TeV brane demanding $-$ parity for the $X,Y$ (and $+$ for the 
$A_\mu^a$) gauge fields.
This deformation in AdS implies in the 4d picture a spontaneous breaking of $SU(5)$ down to the
$SU(3)\times SU(2)\times U(1)$ subgroup at the TeV: the $X,Y$ bosons acquire TeV masses through 
the Higgs mechanism and the CFT resonances are not $SU(5)$ invariant.
At energies higher than the TeV, however, the Planck brane-brane correlator does not 
probe the GUT breaking on the TeV brane and the holographic theory must appear fully $SU(5)$ invariant.
As a consequence, we expect a GUT-invariant running of the $SU(3)\times SU(2)\times U(1)$
gauge couplings $g_i$, $i=1,2,3$, from the TeV up to higher energies. This is indeed what we found 
computing the contribution of the massless 5-plet scalar (for $1/z_1\gg |p|\gg z_0/z_1^2$):
\begin{equation} \label{eq:GUT+-}
\begin{split}
\frac{1}{g_i^2(p^2)} =& \frac{1}{8\pi^2} \log\frac{z_1}{z_0}\, \left[ \frac{8\pi^2}{k g_5^2} - b_5 \right] 
+ \Delta_0(1/z_0) + \Delta_1^i(1/z_1) -\frac{b_5}{8\pi^2} \log z_0\sqrt{-p^2}
  \\
 &- \frac{1}{8\pi^2} 
 \left[ \frac{b_2^i}{2} \left( -\frac{1}{\epsilon} +\frac{\gamma}{2}-\frac{8}{3} \right) -
        \frac{b_3^i}{2} \left( 2\log 2 + \frac{8}{3} \right) \right] \;.
\end{split}
\end{equation}
We denote with $b_{2,3}^i$ the beta-functions of a 4d scalar doublet, triplet respectively and
with $b_5$ the $SU(5)$-invariant beta-function. We recognize in the previous formula (fourth
term) the contribution
of the holographic 5-plet (a massless doublet and a triplet with a tiny mass $\sim\TeV^2/k$),
and the CFT contribution at NLO in $1/N$ (first term). Both are $SU(5)$ invariant as 
expected \footnote{At very low energies, $|p|<z_0/z_1^2$, the triplet contribution stops and
the running becomes different for the three $SU(3)\times SU(2)\times U(1)$ couplings $g_i$.}.
Notice that for this to happen, we had to evaluate the boundary couplings $\Delta_{0,1}(\mu)$
on the Planck and TeV branes at $\mu=k,\TeV$ respectively. This is quite natural as these 
boundary terms $\Delta_{0,1}$ correspond holographically to threshold corrections
at the scales $1/z_0$, $1/z_1$.
Had we evaluated, for instance, the TeV boundary term at $\mu=k$,
a fake $SU(5)$-breaking effect would have been introduced, coming from the $SU(5)$ non-invariant
evolution of $\Delta_1^i(\mu)$. 
The logarithmic divergence is canceled with an $SU(5)$ non-invariant counterterm on the TeV brane: 
the only source of differentiation among the three couplings $g_i$ comes from $\Delta^i_1(1/z_1)$
and from some finite scheme-dependent terms absorbable in $\Delta^i_1$.
Changing the value of the latter, corresponds holographically to modify the Higgs mechanism
responsible for the $SU(5)$ breaking. How much the $g_i$ depart from a common value 
below the TeV depends therefore on the unknown value of $\Delta^i_1(1/z_1)$. 
It is clearly an important phenomenological question to estimate this contribution in some way.
One can advocate a plausible strong coupling hypothesis \cite{Contino:2001si}
assuming that the $\Delta^i_1(\mu)$ are sufficiently small when the gauge dynamics becomes 
strongly coupled. On the TeV brane this happens at energies $\mu\gtrsim \TeV$,
confirming that the choice of the scale $\mu=1/z_1$ for $\Delta_1$ is the correct one.

\vspace{0.3 cm}
\noindent \begin{center} \textsc{GUT breaking on both the TeV and Planck brane} \end{center}
\begin{equation*}
\varphi = \begin{bmatrix} \varphi_2 (++) \\ \varphi_3 (--) \end{bmatrix}
\qquad\qquad\qquad \text{for} \quad A_\mu^{\hat a} (--) \;\; A_5^{\hat a} (++)
\end{equation*}
\vspace{-.1cm}

If $SU(5)$ is broken by the Planck brane boundary conditions, the holographic theory does not
have $X,Y$ bosons. Even if the CFT has a global $SU(5)$ invariance (see section~\ref{brane}),
only the $SU(3)\times SU(2)\times U(1)$ symmetry is gauged. 
In the holographic theory we thus find, in addition to the CFT sector, 
the $SU(3)\times SU(2)\times U(1)$ gauge fields and an elementary doublet scalar.
Inserting the TeV brane in AdS and demanding a $-$ parity for the $A_\mu^{\hat a}$, 
the global $SU(5)$ invariance of the CFT is spontaneously broken at the TeV to the
$SU(3)\times SU(2)\times U(1)$ subgroup. The corresponding Goldstone bosons can be identified with
the zero modes of $A_5^{\hat a}$, which appear in the dual theory as scalar 
excitations of the CFT with the same quantum numbers of the XY bosons.
From the holographic point of view, we thus expect an $SU(5)$-breaking running up to the Planck scale
given by the scalar doublet, while the CFT does not contribute to the differential running. Indeed
the explicit calculation gives:
\begin{equation} \label{eq:GUT--}
\begin{split}
\frac{1}{g_i^2(p^2)} =& \frac{1}{8\pi^2} \log\frac{z_1}{z_0}\, \left[ \frac{8\pi^2}{k g_5^2} - b_5 \right]
+ \Delta^i_0(1/z_0) + \Delta^i_1(1/z_1) -\frac{b_2^i}{8\pi^2} \log z_0\sqrt{-p^2}
 \\
 &- \frac{1}{8\pi^2} 
 \left[ \frac{(b_2^i-b_3^i)}{2} \left( -\frac{1}{\epsilon} +\frac{\gamma}{2} \right) 
        - \frac{b_3^i}{2} \log 2 - b_2^i \frac{4}{3} \right] \;.
\end{split}
\end{equation}

This equation gives a non-ambiguous test of the holographic interpretation: the 4d scalar doublet
gives a differential running up to the scale $k$. This effect cannot be falsified by the 
$SU(5)$-invariant running of the CFT.

An important observation is in order at this point. From the holographic point of view, there is
no reason at all why the different $g_i$ couplings should unify at the Planck scale.
Indeed, in the holographic theory $SU(5)$ is just a global symmetry of the 
pure CFT sector, only the $SU(3)\times SU(2)\times U(1)$ group is gauged.
This is in sharp contrast with the case of $SU(5)$ broken only 
by the TeV brane: in that case, there is a Higgs mechanism in 4d reducing the GUT group at the TeV.
No analogous mechanism arises here at the Planck scale.
Moreover, from the 5d point of view, the situation at energies around $k$ is similar to the
flat case: there is really no exact unification of the gauge couplings just because 
there is no unified symmetry on the boundaries.
As in the flat limit, however, one can estimate the threshold corrections, represented in AdS by
the boundary term $\Delta_0^i(\mu)$, to be small if evaluated at a scale $\mu\sim 1/z_0$
close to the strong dynamics regime. In this sense, we recover an approximate unification
of the couplings $g_i$ at the Planck scale.

\vspace{0.3cm}
\noindent \begin{center} \textsc{GUT breaking on the Planck brane} \end{center}
\begin{equation*}
\varphi = \begin{bmatrix} \varphi_2 (++) \\ \varphi_3 (-+) \end{bmatrix}
\qquad\qquad\qquad \text{for} \quad A_\mu^{\hat a} (-+) \;\; A_5^{\hat a} (+-)
\end{equation*}
\vspace{-.1cm}

The GUT symmetry is still broken on the Planck brane but no more on the TeV, so that the 
holographic picture is much similar to the previous case. Inserting a TeV brane and demanding
a $+$ parity for the $A_\mu^{\hat a}$, it means that $SU(5)$ remains a global symmetry of the CFT:
the CFT resonances can be arranged in exact $SU(5)$ multiplets.
%On the other hand, the $A_5^{\hat a}\to A_5^{\hat a}\, +\,$ const. symmetry
%is violated by the boundary conditions, $A_5^{\hat a}$ being now $(+-)$. Consequently, the
%$\pi$ scalars acquire a mass of order $\sim\TeV$.
As in the previous case we expect that the only source of $SU(5)$ breaking comes from the scalar 
doublet. Indeed we obtain 
\begin{equation}
\begin{split}
\frac{1}{g_i^2(p^2)} =& \frac{1}{8\pi^2} \log\frac{z_1}{z_0}\, 
\left[ \frac{8\pi^2}{k g_5^2} - b_5  \right] + \Delta^i_0(1/z_0) + \Delta_1(1/z_1) 
-\frac{b_2^i}{8\pi^2} \log z_0\sqrt{-p^2} \\
 &- \frac{1}{8\pi^2} 
 \left[ \frac{b_2^i}{2} \left( -\frac{1}{\epsilon} +\frac{\gamma}{2}-\frac{8}{3} \right) +
        \frac{b_3^i}{2} \log 2 \right] \;.
\end{split}
\end{equation}

\subsection{GUT breaking with a bulk vev}

A different mechanism to break the GUT symmetry is the standard Higgs mechanism.
Let us suppose that a massless scalar field $\Sigma$, propagating in the bulk, acquires
a vacuum expectation value $\langle\Sigma\rangle$ constant along the fifth dimension.
In the following we assume that $\Sigma$ and all the other bulk fields have $(++)$
boundary conditions.
This vev splits the masses of the GUT multiplets, giving, for example, a (bulk) mass 
$m\sim g_5\langle\Sigma\rangle$ to the triplet of our scalar $\varphi$, leaving the doublet 
massless. An interesting possibility is that $k\gg m\gg\TeV$ so that the one-loop correction 
to the low energy couplings reads:
\begin{equation} \label{eq:GUTmassive}
\begin{split} 
\frac{1}{g_i^2(p^2)} = & \frac{1}{8\pi^2} \log\frac{z_1}{z_0} \left[ \frac{8\pi^2}{k\, g_5^2} -b_5 
    - b_3^i \frac{m^2 z_0^2}{8} \right] \\ 
 &+ \Delta_0(1/z_0) + \Delta_1(1/z_1)  - \frac{b_2^i}{8\pi^2} \log \frac{\sqrt{-p^2}}{m} 
 -\frac{b_5}{8\pi^2} \log mz_0 \\ 
 &- \frac{1}{8\pi^2} \left[ \frac{b_5}{2}\left(-\frac{1}{\epsilon} +\frac{\gamma}{2}\right)
    - \frac{4}{3}b_2^i - \frac{b_3^i}{2}\log 2 -b_3^i \frac{m^2 z_0^2}{16} \right] \;.
\end{split}
\end{equation}
In the 4d dual picture the gauge symmetry is spontaneously broken. The 4d doublet and triplet scalars  
take different masses (the triplet has a mass $\sim m/\sqrt{2}$, corresponding to the lowest 
eigenvalue of the 5d KK tower): their contribution can be recognized in eq.~(\ref{eq:GUTmassive}).
The vev of the $\Sigma$ field implies, as discussed in section \ref{brane}, that the CFT is not $SU(5)$ invariant. 
Therefore, we expect GUT symmetry breaking terms in the CFT beta-function
proportional to $m^2/k^2$; in fact they appear in the first term of eq.~(\ref{eq:GUTmassive}).
While the $\log m$ can be traced back to a calculable and non-analytic operator 
$(\log\Sigma) FF$ in the AdS effective action,
the $m^2/k^2$ terms come from $\Sigma^2 FF$ operators on the AdS side.
As a last remark, we notice that there are no terms in eq.~(\ref{eq:GUTmassive})
linear in $m$. They would be the counterpart either of an analytic 5d operator $\Sigma F F$ or 
of the non-analytic operator $\sqrt{\Sigma^2}FF$. The first one is absent if we impose a 
$\Sigma \to -\Sigma$ symmetry and the second one shows up only in the flat limit $m \sim 
g_5\langle\Sigma\rangle\gg k$, as already said in section~\ref{low}.

\section{\label{conclusions}Conclusions}

We have studied the dynamics of gauge interactions in the Randall-Sundrum model with gauge bosons in 
the bulk, which is conjectured to be dual to a 4d CFT weakly coupled to the corresponding 4d gauge sector.
This duality allows us to keep a perturbative control on the model up to Planck scale, if we limit our study to 
inclusive correlators on the Planck brane.
The evolution of the gauge couplings up to high energies in the holographic theory
gives an insight of the dynamics in the 5d theory.
Bulk loop corrections to brane-brane correlators give both the 
$1/N$ expansion of the CFT and the ordinary perturbative expansion in powers of the gauge coupling constant. 

Using dimensional regularization, we have calculated the 
1-loop correction to the low-energy gauge couplings in 5d due to a bulk scalar
with various boundary conditions on the two branes 
and arbitrary mass. These zero-mode propagators give the 4d holographic couplings at low energy with
their evolution from the Planck scale.

The calculations allowed us to study different GUT scenarios where 
the gauge symmetry is broken either by a Higgs mechanism, or by the boundary conditions. 
We have checked that in any case the results are compatible with what expected 
from the holographic dual. 

Some general conclusions can be drawn for model building. We have seen that, as the CFT has a positive 
beta-function, strong limits are obtained if one imposes that the gauge coupling remains perturbative
up to a standard GUT scale ($\sim 10^{16}\;\GeV$): roughly speaking, the CFT has not to be dominant with respect
to the other contributions, so that large values of $N$ are forbidden. This in turn implies that 
subleading CFT contribution is typically negligible. 

Different phenomenological models are possible.
If the Standard Model particles are confined on the Planck brane, supersymmetry is required to stabilize
the hierarchy; one reobtains a standard 
supersymmetric unification, if a spontaneous breaking occurs on the Planck brane \cite{Pomarol:2000hp}.
From the 4d point of view, we have just added to the MSSM a GUT-invariant CFT, which just gives a common 
positive contribution to all the three beta-functions. 

As discussed in \cite{Randall:2001gb}, we can also
imagine to put the Standard Model on the TeV brane, in order to solve the hierarchy problem.
In this case, proton decay mediated by X,Y Kaluza-Klein bosons with TeV masses must be forbidden;
for example by choosing Dirichlet boundary conditions for the broken
gauge bosons and requiring additional symmetries for the TeV brane interactions.
This breaking of the GUT symmetry through TeV brane boundary conditions is negligible 
for energies above the TeV scale; additional sources of GUT breaking are therefore 
required, such as a Higgs mechanism in the bulk or on the Planck brane.

If the only source of symmetry breaking is the choice of boundary conditions on the TeV brane, the
unification scale should be at the TeV scale. This could fit well in the framework of $SU(3)_W$ unification 
\cite{Weinberg:1971nd}, recently readdressed in extra-dimensional inspired models 
\cite{Dimopoulos:2002mv}, in which the $SU(2)$ and $U(1)$ groups of the Standard Model 
are embedded into a weak $SU(3)$ around the TeV scale. 
However, it is likely that this kind of model requires a scale of conformal symmetry breaking
too low to be compatible with the strong limits coming from electroweak precision observables 
\cite{Csaki:2002gy}.

A further possibility is the breaking through Planck brane boundary conditions.
In this case, there is no unification in the usual sense, as only the SM gauge bosons exist in the 
holographic dual .
Nevertheless, as in the flat case, an approximate unification at high energies can be justified from a 
5d point of view, relying on a strong coupling assumption for the boundary couplings on the Planck brane.

Using both the AdS picture and the 4d dual counterpart, unification of gauge couplings in these warped 
spaces can be discussed. Only further work will tell us if a viable and compelling model is achievable.

%%%%%%%%%%
%%%%%%%%%%     Acknowledgments
%%%%%%%%%%

\section*{Acknowledgments}

We would like to thank R. Sundrum for useful discussions.
We are especially grateful to R. Rattazzi who has followed this work from the beginning
with many important discussions and suggestions. 
R.C. and E.T. thank the CERN Theory Division, where part of this work was done, for its hospitality.
R.C. also acknowledges the hospitality and the financial support of the Department of Physics 
at the University of Geneve.
This work was partially supported by the EC under TMR contract HPRN-CT-2000-00148.  
\vspace{-0.1cm}

%%%%%%%%%%
%%%%%%%%%%     Appendix
%%%%%%%%%%

\appendix
\section*{Appendix}
\section{Sums in AdS}

We present here the method used to sum the series of eq.~(\ref{eq:pisum}) on the 
AdS Kaluza-Klein masses.
Performing first the integral in eq.~(\ref{eq:pisum}), one find the series
\begin{equation} \label{eq:series}
S(d) = \sum_{\{x_n\}} \left(x_n^2+c^2(x)\right)^{d/2-2} \;,
\end{equation}
where the summation runs over the entire KK spectrum of the scalar field.
Depending on its boundary conditions, the KK masses $x_n$ of a massive scalar field in AdS
satisfy the following eigenvalue equations:
\begin{equation}
\label{eq:spectrum}
\begin{aligned}
(++): \qquad \frac{j_\nu(x_n z_0)}{y_\nu(x_n z_0)} &= \frac{j_\nu(x_n z_1)}{y_\nu(x_n z_1)}\, ; \\
(+-): \qquad \frac{j_\nu(x_n z_0)}{y_\nu(x_n z_0)} &= \frac{J_\nu(x_n z_1)}{Y_\nu(x_n z_1)}\, ;
\end{aligned} \qquad\qquad
\begin{aligned}
(--): \qquad \frac{J_\nu(x_n z_0)}{Y_\nu(x_n z_0)} &= \frac{J_\nu(x_n z_1)}{Y_\nu(x_n z_1)} \\   
(-+): \qquad \frac{J_\nu(x_n z_0)}{Y_\nu(x_n z_0)} &= \frac{j_\nu(x_n z_1)}{y_\nu(x_n z_1)} 
\end{aligned}
\end{equation}
Here $J_\nu$, $Y_\nu$ are Bessel functions, $\nu=\sqrt{4+m^2 z_0^2}$, with $m$ the 5d mass, and
\begin{equation}
y_\nu(z) = Y_{\nu-1}(z) + \frac{(2-\nu)}{z} Y_\nu(z)\, ; \qquad 
j_\nu(z) = J_{\nu-1}(z) + \frac{(2-\nu)}{z} J_\nu(z) \;.
\end{equation}
Choosing the functions:
\begin{equation} \label{eq:functions}
\begin{split}
f_{++}(z) &= y_\nu(z z_1) j_\nu(zz_0) - y_\nu(z z_0) j_\nu(zz_1) \\
f_{--}(z) &= Y_\nu(z z_1) J_\nu(zz_0) - Y_\nu(z z_0) J_\nu(zz_1) \\
f_{+-}(z) &= i \left[ y_\nu(z z_0) J_\nu(zz_1) - Y_\nu(z z_1) j_\nu(zz_0) \right] \\
f_{-+}(z) &= i \left[ y_\nu(z z_1) J_\nu(zz_0) - Y_\nu(z z_0) j_\nu(zz_1) \right] \;,
\end{split}
\end{equation}
whose zeros are the $x_n$s, one can rewrite the sum in eq.~(\ref{eq:series}) 
as a complex integral over the contour $\Gamma$ with $R\to\infty$ 
(see fig.~\ref{contour}):
\begin{equation}
S(d) = \frac{1}{2\pi i} \int_\Gamma \!\! dz\, \left(z^2+c^2\right)^{d/2-2}\, \frac{f'(z)}{f(z)}
\end{equation}
with $f$ one of the functions in eq.~(\ref{eq:functions}) \footnote{Even if the domain of 
definition of the Bessel functions $J_\nu(z)$, $Y_\nu(z)$ is the $z$-plane cut along the 
negative real axis, the functions $f_{\pm,\pm}(z)$, $f_{\pm\mp}(z)$ are single-valued on 
the entire complex plane.}.
\begin{figure}\begin{center}
\begin{picture}(200,140)
\Line(20,-10)(20,30) \LongArrow(20,110)(20,150) 
\ZigZag(20,30)(20,110){1.3}{16} \LongArrow(0,70)(160,70)
\Vertex(20,110){1.8} \Vertex(20,30){1.8} \Text(38,64)[]{$\varepsilon$}
\Text(5,143)[]{\small Im$\, z$} \Text(150,62)[]{\small Re$\, z$}
\Text(10,110)[r]{$ic$} \Text(10,30)[r]{$-ic$}
\ArrowLine(30,140)(30,70) \ArrowLine(30,70)(30,0) 
\ArrowArc(20,70)(70.7,-81,0) \CArc(20,70)(70.7,0,81) \Text(90,30)[]{$\Gamma$}
\LongArrow(20,69)(70,119) \Text(55,93)[]{$R$}
\Text(55,70)[]{$\scriptstyle \times$} \Text(70,70)[]{$\scriptstyle \times$}
\Text(85,70)[]{$\scriptstyle \times$} \Text(100,70)[]{$\scriptstyle \times$}
\Text(115,70)[]{$\scriptstyle \times$} \Text(130,70)[]{$\scriptstyle \times$}
\end{picture} \vspace{0.5cm}
\caption{\label{contour} \small Contour $\Gamma$ in the complex plane. 
The crosses along the real axis correspond to the
real positive zeros $x_n$ of the function $f$.}
\end{center} \end{figure}
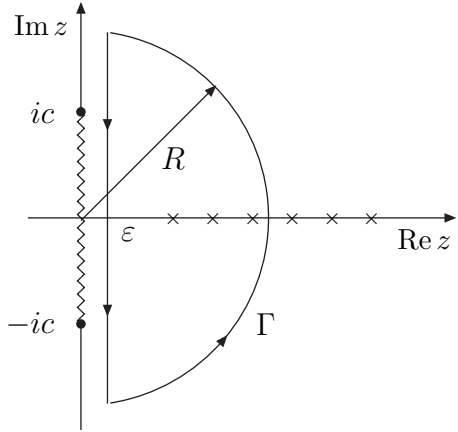
What follows applies for a generic parity, and therefore we will specify the function
$f$ only when necessary.
The asymptotic expansion of $f(z)$ when $\text{Im}\, z\to\pm\infty$, the same for all the parities,
\begin{equation}
\frac{f'(z)}{f(z)} = -\left( \pm i(z_1-z_0) + \frac{1}{z} \right) + O(1/z^2) 
\end{equation}
tells us that the integral, like the original series, converges at infinity ($R\to\infty$) if $d<3$.
In order to find the expression of $S(d)$ for $d\to 4$, we first take $d<3$ and extract the limit 
$R\to\infty$. The contribution of the integration around the semi-circle of radius $R$ goes to zero
and we are left with the vertical contour. Let us call for convenience $\Gamma^+$, $\Gamma^-$ 
the part of this vertical contour respectively above, below the real axis.
We now subtract the asymptotic behaviour of $f'/f$ and evaluate it separately 
deforming $\Gamma^+$ and $\Gamma^-$ to coincide with the real axis. Defining
\begin{equation}
F(z) = \frac{f'(z)}{f(z)} + \frac{1}{z} +i(z_1-z_0)
\end{equation}
and using the parity properties $f_{\pm\pm}(-z)=f_{\pm\pm}(z)$, $f_{\pm\mp}(-z)=-f_{\pm\mp}(z)$,
we obtain:
\begin{equation}
\begin{split}
S(d) =& \frac{1}{2\pi i} \left[ \int_{\Gamma^+} \!\! dz\, \left(z^2+c^2\right)^{d/2-2} F(z) -
 \int_{\Gamma^-} \!\! dz\, \left(z^2+c^2\right)^{d/2-2} F(-z) \right] \\
 &+ \frac{(z_1-z_0)}{2\sqrt{\pi}} \left(c^2\right)^{(d-3)/2} 
    \frac{\Gamma\left(\frac{3-d}{2}\right)}{\Gamma(2-d/2)} \;.
\end{split}
\end{equation}
In the remaining integrals, we can now extract the limit $d\to 4$, being $F(z)\sim 1/z^2$ at infinity.
We expand the integrand up to orders $O[(d-4)^2]$. The first term in the expansion gives a non-vanishing
result because of a residue contribution in the origin (here we deform the contours $\Gamma^+$, $\Gamma^-$
to coincide with the imaginary axis, $\varepsilon\to 0$ in fig.~\ref{contour}):
\begin{equation}
\frac{1}{2\pi i} \left[ \int_{\Gamma^+} \!\! dz\, F(z) - \int_{\Gamma^-} \!\! dz\, F(-z) \right]
 = \frac{\alpha}{2}\, ; \qquad \quad
 \alpha = \begin{cases} \pm 1 &\text{for}\; (\pm,\pm) \\ 0 &\text{for}\; (\pm,\mp) \end{cases} \;. 
\end{equation}
The second term in the expansion must be evaluated taking into account the cut along the imaginary axis
between $\pm ic$. We find:
\begin{equation}
\begin{split}
\frac{1}{2\pi i} \bigg[ \int_{\Gamma^+} \!\! dz\, \log\left(z^2+c^2\right) F(z) 
 - &\int_{\Gamma^-} \!\! dz\, \log \left(z^2+c^2\right) F(-z) \bigg] \\
 &=\log f(ic) + \log c\pi\sqrt{z_0 z_1} + \alpha \log c -c (z_1-z_0) \;.
\end{split}
\end{equation}
Summing all the contributions we get our final result
\begin{equation}
\sum_{\{x_n\}} \left(x_n^2+c^2(x)\right)^{d/2-2} = \frac{\alpha}{2} + (d/2-2)
 \Big[ \log f(ic) + \log c\pi\sqrt{z_0 z_1} + \alpha \log c \Big] + O[(d-4)^2]
\end{equation}
which leads to eq.~(\ref{eq:pi}).
Finally, we write the $z\to 0$ expansion of the various functions $f$, used in the text to obtain
the low energy limit of $\Pi(p^2,\mu)$. Taking only the relevant terms, one has ($z\to 0$, $z_1 \gg z_0$):
\begin{equation}
\label{eq:asymptotes}
\begin{aligned}
f_{++}(z) &\simeq \frac{1}{\pi\nu} \left(\frac{z_1}{z_0}\right)^{\nu-1} 
 \left[ \frac{4-\nu^2}{z^2 z_0^2} + \frac{2+\nu}{2(\nu-1)} \right] \\
f_{--}(z) &\simeq \frac{1}{\pi\nu} \left(\frac{z_1}{z_0}\right)^{\nu} \\ 
f_{+-}(z) &\simeq \frac{i}{\pi\nu} \left(\frac{z_1}{z_0}\right)^{\nu} 
 \left[ \frac{\nu-2}{zz_0} - \frac{zz_0}{2(\nu-1)} + \frac{\nu+2}{zz_0} 
        \left(\frac{z_0}{z_1}\right)^{2\nu} \right] \\
f_{-+}(z) &\simeq \frac{i}{\pi\nu} \left(\frac{z_1}{z_0}\right)^{\nu} \frac{2+\nu}{zz_1} \;.
\end{aligned}
\end{equation}
%

%%%%%%%%%%
%%%%%%%%%%    References
%%%%%%%%%%

\end{document}